\documentclass[a4paper]{article}
\usepackage{Odyssey2022}
\usepackage{epsfig,amssymb,amsmath,color,xcolor}
\ninept
\usepackage{url}

\title{Gamified Speaker Comparison by Listening}

\makeatletter
\def\name#1{\gdef\@name{#1\\}}
\makeatother
\name{{\em Sandip Ghimire$^1$, Tomi Kinnunen$^1$, Rosa Gonz\'alez Hautamäki$^2$}}

\address{$^1$School of Computing, University of Eastern Finland, Finland\\
$^2$Department of Electrical \& Computer Engineering,
National University of Singapore, Singapore}
\begin{document}
\maketitle

\begin{abstract}
We address speaker comparison by listening in a game-like environment, hypothesized to make the task more motivating for naive listeners. We present the same 30 trials selected with the help of an x-vector speaker recognition system from VoxCeleb to a total of 150 crowdworkers recruited through Amazon's Mechanical Turk. They are divided into cohorts of 50, each using one of three alternative interface designs: (i) a traditional (non-gamified) design; (ii) a gamified design with feedback on decisions, along with points, game level indications, and possibility for interface customization; (iii) another gamified design with an additional constraint of maximum of 5 `lives' consumed by wrong answers. We analyze the impact of these interface designs to listener error rates (both misses and false alarms), probability calibration, time of quitting, along with survey questionnaire. The results indicate improved performance from (i) to (ii) and (iii), particularly in terms of balancing the two types of detection errors. 
\end{abstract}

\section{Introduction}

\emph{Speaker comparison} involves comparing two speech samples to decide whether or not they were produced by the same individual. It can be done by humans through the process of listening \cite{SchmidtNielsen2000,Shen2011}, or by computers by leveraging from signal processing and machine learning techniques. 
We focus on the former.

Even if automatic speaker recognition is scalable and often surpasses humans, 
speaker comparison by listening does have its applications. First, manual or semi-automatic \emph{auditing} \cite{CallMyNet2_JonesSWW20,DoddingtonPMR00,Chodroff2017-mixer6} of speaker and language labels in collected speech corpora is often necessary for quality reassurance, especially for automatically harvested data. 
Second, performance assessment of text-to-speech and voice conversion systems in their ability to match or clone individual properties of voice also relies on listening \cite{Zhao2020-vcc2020}.
Finally, given that listeners and automatic systems might use different traits in speaker comparison, the two could potentially be combined in \emph{human-in-the-loop} systems. 


Unfortunately, speaker comparison is not a natural task for listeners. While we may recognize \emph{familiar} speakers (such as family members) \cite{maguinness2018understanding} accurately,
%
the above applications more often involve speakers unfamiliar to the listener. Thus, such task might be cognitively demanding or boring, with the sole motivation of a listener obtaining a monetary or other kind of external reward. As a result, the performance of listeners might not reach its full potential. 

\begin{figure}[!t]
\centering
\includegraphics[scale=0.4]{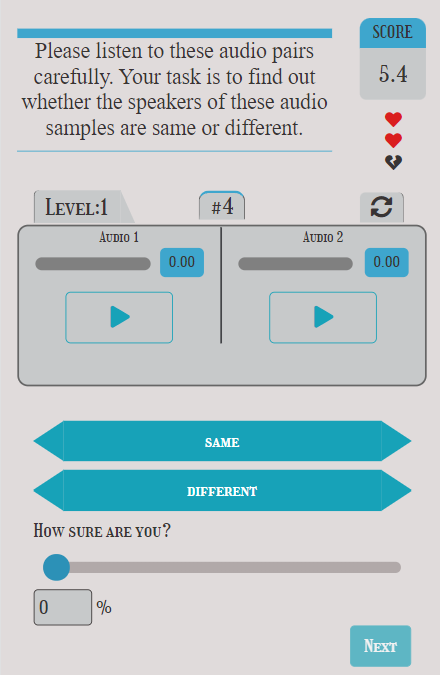}\\
\caption{Gamified speaker comparison framework}
\label{fig:game_design}
\end{figure}

In this study, we address the question whether speaker comparison could be made more interesting or rewarding to listeners, by leveraging from a game-like environment (see Fig. \ref{fig:game_design}). \emph{Gamification} \cite{deterding2011} refers to use of game-inspired design, or game elements used in non-gaming contexts. Compared to conventional `passive' tasks, gamification enables goal-directed behaviour by introducing concepts such as points, badges, difficulty levels or leaderboards. As such, gamification has the potential to increase user engagement and motivation for performing a given task \cite{sailer2016,hamari2017,jared2018}. 
While gamification has been used in domains ranging from crowdsourcing and data-collection to health, marketing and social networks, there is less work in the speech domain. 
We are not aware of prior work in gamified speaker comparison by listening. Therefore, as a pilot study of its kind, our interest is to address a number of basic questions, including: 
    \begin{enumerate}
        \item Can gamification improve speaker detection performance of listeners?
        \item How does gamification impact the two different types of detection error rates (miss and false alarm rate)?
        \item How does gamification impact confidence of listeners?
        \item How does gamification impact listener's opinion of the meaningfulness of the task?
    \end{enumerate}
To address these questions, we designed a gamified speaker comparison framework 
customized to create three different interfaces, one traditional speaker comparison interface and two gamified interfaces \footnote{Source code available at \url{https://github.com/sandip-ghimire/VoiceComp}}. Through the use of otherwise shared listening trials, but varied interface designs, we can address the impact to a number of objective and subjective measures defined below. Using Amazon's Mechanical Turk (AMT) service, we recruited 50 crowdworkers per each of the three interface designs. The trials (maximum 30 per scenario) are sampled from VoxCeleb1 \cite{nagrani17}. The following sections provide further background, details of our study, followed by results.

\section{Background}\label{sec:speech-gamification-review}

\subsection{Crowdsourcing}

\emph{Crowdsourcing} is one of the attractive target domains for integrating gamification for empirical data-collection. Crowdsourcing outsources a given tasks to a large group of individuals known as \emph{(crowd)workers}. \cite{howe2006,liu2014}. 
The workers are rewarded if they fulfill the requirements set by the data collector. Compared to traditional data collection in a laboratory environment, crowdsourcing enables collecting large amounts of data from diverse populations in short turnaround time. In the speech field, examples of crowdsourcing include performance assessment of naive listeners in speaker comparison \cite{Shen2011-mechanical-turk}, speech perception studies \cite{cooke2013} and collecting mean opinion scores for subjective assessment of synthetic speech quality \cite{Cooper21-how-do-voices} to name a few.



Nonetheless, the typical crowdsourcing designs are centered around the requester (experimenter) whose motivations are not necessarily aligned with those of the workers. In general, motivations can be divided into \emph{intrinsic} and \emph{extrinsic} ones \cite{ryan2000}. The former refers to genuine interest in (or enjoyment of) an action for its own sake and reflects our abilities to learn and assimilate. The latter, in turn, refers to motivations driven by external rewards such as money, praise, or certificate \cite{odobasic2013}. Commercial crowdsourcing platforms provide way to connect workers and requesters. 
While money is a clear extrinsic motivator, we hypothesize that we could do better by gamification. We hypothesize that motivated workers could potentially be more accurate or more cost-efficient (less errors in given time-frame). 

\subsection{Gamification}

The design and aesthetics play key roles in designing gamified tasks. One framework for game design is MDA \emph{(mechanics, dynamics, aesthetics)} \cite{hunicke2004}. Mechanics describe the components of the game, dynamics describe the run-time behavior and aesthetics detail the emotional responses evoked in players. These components encompass game design elements such as rewards, stories, feedback and progress. 

\textbf{Points} are the basic elements of games and gamified environments. They are the reward provided for successfully completing a task or activities in a gamified environment \cite{werbach2012}. Point can also be used to give the feedback or to numerically represent a player's progress in the game \cite{sailer2016}. Points serve as measurable performance indicator which instantly indicates how the user performs in the game. \textbf{Levels} symbolize the completion of certain steps in the game and represents the achievement in a cumulative way. Levels also represent how close the player is towards the goal, which may help to encourage or motivate a player. It may represent the virtual achievement to help boost the confidence of the player. Finally, \textbf{stories} give the narrative context and instructions for conveying information and directions to the user. It can be communicated through simple instructions or storylines of contemporary role-playing \cite{kapp2012}. Narrative contexts can be designed to simulate the real-world context or relay the idea of non-gaming scenarios in gaming context which might excite the player and motivate in otherwise boring tasks \cite{reiners2014}.



\subsection{Game-like elements in speaker and language recognition}

Even if our study focuses on speaker comparison by listening, we also identify a number of game-like environments used in the automatic speaker and language recognition research. As an example, some of the NIST SRE campaigns (including 2021 CTS challenge \cite{Sadjadi2021}) include periodical \emph{leaderboard} snapshot on the ranking of different teams or individuals. 

Concerning speaker comparison by listening, \emph{stories} and \emph{role plays} have been used in a number of studies. For instance, \cite{Wu2016-sas-corpus} instructs listeners to imagine they work in a bank and in \cite{rosa2020} we addressed the impact of different role play framings to miss and false alarm rates. In \cite{Shen2011-mechanical-turk} the authors made payment to crowdworkers \emph{conditional} on their speaker recognition performance, and included multiple-choice questions on speech content to motivate the workers to actually listen to the two audio samples in question. 

Recently, \cite{cieri2021} used a language recognition game to help data labeling effort. In particular, the authors published `\emph{Name That Language!}' (NTL) tool to help build corpora for language recognition. The players listen to audio clips of known or suspected languages and indicate the language spoken.  The listener receives 10 points per correct answer or lose one of three lives for incorrect ones. The difficulty level increases as the listener advances in the game, where no feedback is provided but points are given for each answer. 


\section{Experimental Setup}

\subsection{Speaker comparison with difficulty levels}

To assess the potential benefits of gamification in speaker comparison, we set-up a speaker comparison interface where crowdworkers listen to pairs of speech samples and submit binary decisions (same/different speaker) along with \emph{confidence} of the decision --- a slider with range 0\%--100\%. We include a total of 30 trials, organized into three difficulty levels. Each level consists of ten trials balanced across the two classes (5 target and 5 nontarget trials). This information (balanced class prior) is not informed to the listeners.

The trials were selected from VoxCeleb1 \cite{nagrani17}\footnote{\url{https://www.robots.ox.ac.uk/~vgg/data/voxceleb/vox1.html}}. The three difficulty levels were determined with the aid of an automatic speaker verification system (ASV) with the assumption that the trial difficulties for modern ASV and average listener are sufficiently correlated. To this end, we use an \emph{x-vector} speaker embedding system \cite{snyder2018} with probabilistic linear discriminant analysis (PLDA) back-end. We first sort the scores (separately for targets and nontargets), followed by partition into three equally-sized bins. For the first level, five target trials are randomly selected from the bin with highest scores; similarly, we include five non-target trials selected from the \emph{lowest} scores. We then proceed similarly for the two successive levels of increasing difficulty. The final level contains trials with abnormally high non-target scores and abnormally low target scores.

\begin{table*}[!t]
\caption{Averaged per-listener error rates computed from all 30 trials. The values following $\pm$ indicate the 90\% standard error of the mean (SEM).}
\centering
\begin{tabular}{lccc}
 & $P_\text{miss}$ (\%) & $P_\text{fa}$ (\%) & $P_\text{e}$ (\%) \\ \hline\hline
\texttt{NON-GAMIFIED} & $38.52 \pm 4.56 $ & $10.69 \pm 3.92$ & $25.42 \pm 2.99$ \\ \hline
\texttt{GAMIFIED-UNLIMITED} & $29.31 \pm 3.36$ & $13.26 \pm 1.95$ & $21.65 \pm 2.43$ \\ \hline
\texttt{GAMIFIED-LIMITED} & $34.48 \pm 4,26$ & $12.59 \pm 2.80$ & $24.06 \pm 2.04$ \\ \hline
\end{tabular}
\label{tab:average_error_rates}
\end{table*}

The pre-selected 30 trials are fixed throughout our experiments; what we vary are the game interfaces 
but the speech data remains the same. 
Within each difficulty level, the order of trials is randomized per worker. 

\subsection{The three different interface designs}

We consider three different kinds of interfaces: 
    \begin{enumerate}
        \item \texttt{NON-GAMIFIED}: Non-gamified (traditional) speaker comparison reveals neither the difficulty levels nor provides feedback on the correctness of answers. The user interface is fixed (no possibility for customization).
        \item \texttt{GAMIFIED-UNLIMITED}: Gamification with unlimited lives displays the game level; a score; both visual and aural (sound effect) feedback on the correctness of response; and allows user to customize the interface theme (light or dark theme).
        \item \texttt{GAMIFIED-LIMITED}: As the above but with a constraint on number of `lives', displayed to the user. Once the maximum number of lives has been consumed, `game over' takes place.
    \end{enumerate}
We expect performance difference in speaker discrimination between the first two designs due to the feedback provided; gamified design with feedback can be viewed as a form of \emph{familiarization phase} known to help subjects in adjusting their strategy \cite{bech2007perceptual}.

\subsection{Crowdworkers: Recruitment and Rewards}

In each of the above designs, each worker is required to complete the first 10 trials (the first level). After this mandatory part, the worker may continue or quit at any time. In the first two interface designs, therefore, the worker can decide to complete all the 30 trials, whereas in \texttt{GAMIFIED-LIMITED} this will depend both on the listener's speaker discrimination ability as well as the maximum number of lives.

We use Amazon's Mechanical Turk (AMT) service for crowdworker recruitment, where each trial can be considered as a \emph{Human Intelligent Task} (HIT). Similar to keeping the same speech material in the three different interface designs, we fix the (maximum) reward given to each worker, regardless of the interface they use or what their performance is. In specific, we pay $0.10$ USD for completing each trial along with an additional bonus of $0.50$ USD for completing all the 30 trials. Therefore, each worker gets a minimum reward of $1.00$ USD and maximum of $3.5$ USD.

We recruited 50 workers to use each of the three interface design (a total of 150 workers). They were provided with an introductory consent paragraph with basic information about the 
the experiment along with contact for further information and reassurance that their crowdworker ID would not be made public.
The instructions about the task, reward and quitting options were provided before the link to the game was provided. We did not collect any demographic information of the listeners, though we instructed them to proceed only if they do not have known hearing impairments. The participation was limited  to crowdworkers with the geographical location of United States and crowdwork approval rate of at least $98$\%. The crowdworkers' completed HITs were approved within two hours after completion.

\begin{table*}[ht!]
\caption{Per-listener averaged error rates broken down across the three difficulty levels. The values following $\pm$ indicate the 90 \% standard error of the mean (SEM). The increasing uncertainty of error rates as function of the level in \texttt{GAMIFIED-LIMITED} is explained by the substantially reduced data as workers `die' before reaching the more challenging levels.}
\centering
\begin{tabular}{llcccc}
\multicolumn{ 1}{l}{} & Level & $P_\text{miss}$ & $P_\text{fa}$ & \multicolumn{1}{c}{$P_\text{e}$} & $P_\text{miss}/ P_\text{fa}$\\ \hline\hline
\multicolumn{ 1}{c}{\texttt{NON-GAMIFIED}} & L1 & $23.40 \pm 4.69$ & $8.40 \pm 3.65$ & $16.00 \pm 3.29$ & $2.78$\\ 
\multicolumn{ 1}{l}{} & L2 & $63.06 \pm 4.84$ & $13.64 \pm 7.62$ & $41.52 \pm 5.21$ & $4.62$ \\ 
\multicolumn{ 1}{l}{} & L3 & $60.77 \pm 5.81$ & $10.77 \pm 5.23$ & $35.51 \pm 3.16$ & $5.64$ \\ \hline
\multicolumn{ 1}{c}{\texttt{GAMIFIED-UNLIMITED}} & L1 & $16.00 \pm 4.10$ & $6.40 \pm 2.57$ & $11.20 \pm 2.57$ & $2.50$\\ 
\multicolumn{ 1}{l}{} & L2 & $47.84 \pm 6.28$ & $21.32 \pm 5.53 $ & $37.16 \pm 4.19$& $2.24$ \\ 
\multicolumn{ 1}{l}{} & L3 & $37.78 \pm 5.95$ & $25.19 \pm 6.71$ & $31.48 \pm 4.70$ & $1.49$ \\ \hline
\multicolumn{ 1}{c}{\texttt{GAMIFIED-LIMITED}} & L1 & $23.40 \pm 5.40$ & $5.60 \pm 2.50$ & $14.55 \pm 2.85$ & $4.17$ \\ 
\multicolumn{ 1}{l}{} & L2 & $51.53 \pm 7.27$ & $27.86 \pm 7.44$ & $42.47 \pm 6.20$ & $1.05$ \\ 
 & L3 & $61.11 \pm 17.01$ & $58.10 \pm 24.95$ & $70.11 \pm 20.28$  & $0.88$\\ \hline
\end{tabular}
\label{tab:error-rates-by-level}
\end{table*}

\begin{figure*}[!t]
\centering
\includegraphics[scale=0.45]{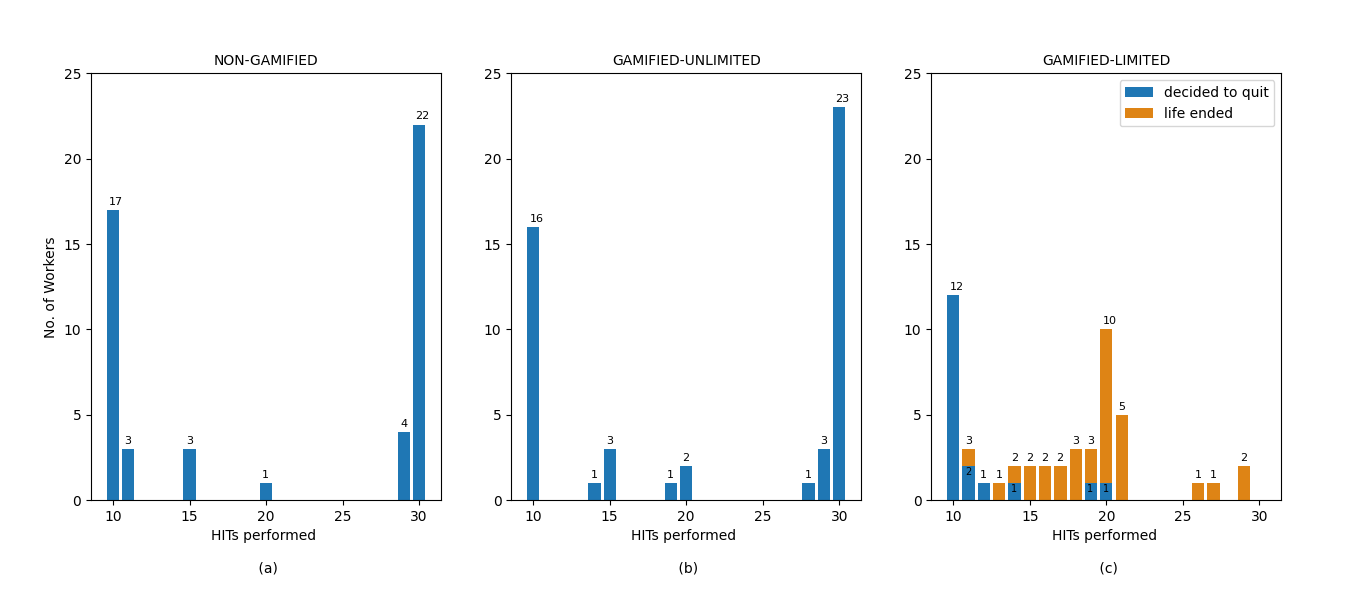}\\
\caption{Number of workers quitting the survey at individual HITs level. HIT refers to a trial.}
\label{fig:quit_chart}
\end{figure*}

\subsection{Interface implementation}

The speaker comparison interface is a web application that can be accessed through a URL\footnote{\url{https://www.voicecomp.net/}} from any web browser. The application runs independently from the AMT service and is accessed by the worker through the URL. The worker is identified by a unique token generated automatically by the application. The application was designed to support several devices and browsers. In order to make the application accessible across multiple platforms, a \emph{responsive design} approach has been adopted keeping aesthetic and functionality intact. The major technologies involved for the application development are HTML, CSS, JavaScript, Bootstrap and jQuery for front-end interface and Python for back-end processes.

\subsection{Data being collected}

We gather several numerical measures from each worker. For each trial, the worker indicates both a binary decision and a confidence value. With the knowledge of the trial ground truth we can objectively assess the speaker discrimination capability of the worker. It is important to keep in mind that after completing the first level, the worker is allowed to quit any time, or might drop out after consuming all `lives' in \texttt{GAMIFIED-LIMITED}. This leads to varied numbers of trials completed. Let $N_\text{tar}(i)$ and $N_\text{non}(i)$ denote, respectively, the total number of target and nontarget trials completed by worker $i$. Likewise, let $N_\text{miss}(i)$ and $N_\text{fa}(i)$ be the number of \emph{misses} (worker declares a same-speaker trial as different-speaker trial) and \emph{false alarms} (worker declares a different-speaker trial as same-speaker trial) by worker $i$. The per-worker empirical \textbf{miss rate}, \textbf{false alarm rate} and \textbf{total error rate} are then computed as the fractions
    \begin{equation}
        \begin{aligned}
            P_\text{miss}(i) & = \frac{N_\text{miss}(i)}{N_\text{tar}(i)}\; \times 100\% \nonumber\\
            P_\text{fa}(i) & = \frac{N_\text{fa}(i)}{N_\text{non}(i)}\; \times 100\% \nonumber\\
            P_\text{e}(i) & = \frac{N_\text{miss}(i)+N_\text{fa}(i)}{N_\text{tar}(i)+N_\text{non}(i)} \; \times 100\%. \nonumber\\
        \end{aligned}
    \end{equation}
For comparisons where the interest is to contrast different interface designs we consider all the 50 workers in a given interface design in statistical terms. To this end, we report the average of the above per-worker quantities, along with the 90\% standard error of the mean (SEM).

Besides the speaker discrimination related measures, we collect information on the time used by the worker on each trial along with an optional survey with 5-point Likert scale response options (detailed below).


\section{Results}

\subsection{Error rates}

As our first analysis, Table \ref{tab:average_error_rates} displays the averaged per-worker miss, false alarm and total error rates for the different interface designs and all trials. 
Focusing first on the total error rate, both of the gamified designs yield less errors (on average) than the traditional (non-gamified) set-up. Increased listener performance was expected, given that the gamified designs allow feedback and possibility to adjust listening strategies. 

Another interesting finding concerns both the absolute and relative levels of miss and false alarm rates. As for the \emph{absolute} levels, the mean false alarm rate falls on the narrowish range $\sim 10.7\%\dots 13.3\%$ while the miss rates are roughly three-fold, $\sim 29.3\%\dots 38.5\%$. In average terms, our workers were therefore far better in detecting `impostors' than the presence of same person in the two utterances. 

Concerning \emph{relative} levels of miss and false alarm rates, we observe a trade-off when comparing any pair of the three interface designs: whenever the miss rate decreases, false alarm rate increases (and vice versa). Such error trading off, achievable by adjusting a detection threshold (operating point) is well-known by the speaker recognition community. Even though we must avoid drawing such analogy too far (remembering Table \ref{tab:average_error_rates} does not present a single `system' but average over many heterogenous `systems'), it is instructive to observe this general pattern. Comparing \texttt{NON-GAMIFIED} and \texttt{GAMIFIED-UNLIMITED}, we observe a substantial decrease in the miss rate (with increase in false alarm rate). Between the two gamified set-ups, the version with limited lives trades reduction in the false alarm rate with an increase in the miss rate.


\begin{figure*}[!t]
\centering
\includegraphics[scale=0.45]{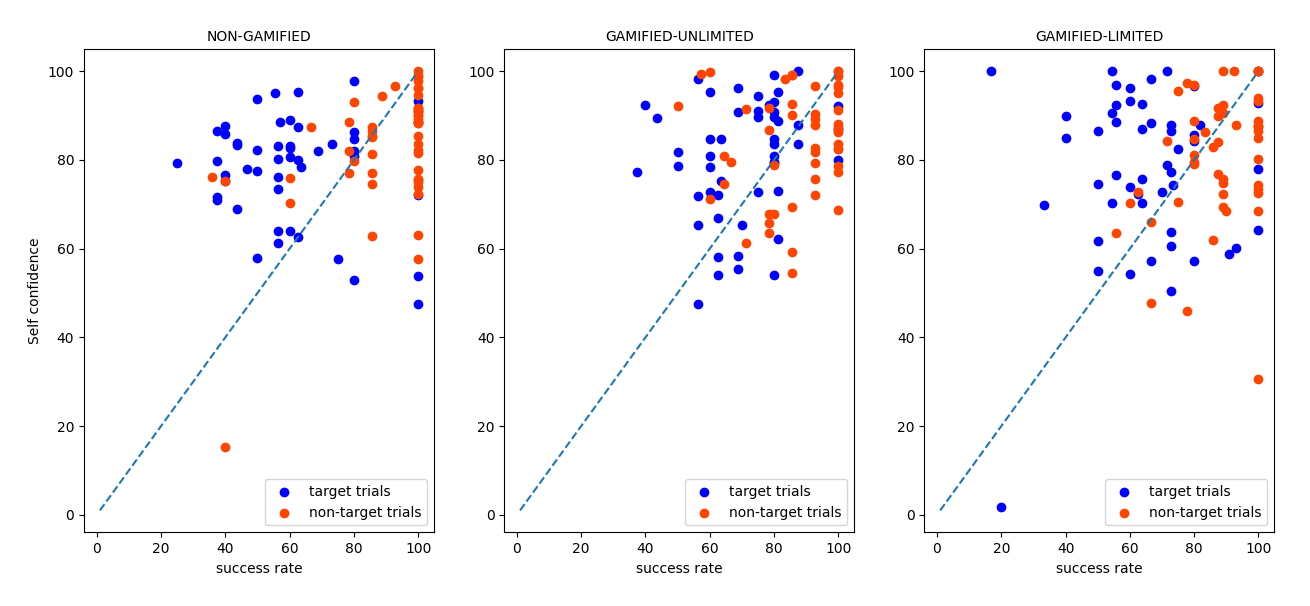}\\
\caption{Calibration graph showing success rate for target and non-target trials against self-reported confidence level}
\label{fig:confidence_plot}
\end{figure*}

Table \ref{tab:error-rates-by-level} shows the averaged per-listener error rates broken down further according to the difficulty level. As expected, the first level is the easiest regardless of the interface design. By treating this mandatory part a common familiarization phase (albeit without feedback in the non-gamified design), let us focus to differences 
at levels 2 and 3. 

Concerning level 2, when moving from the non-gamified design to either gamified design, the average miss rates are decreased --- again with increase in the false alarm rate. Observations concerning level 3 are similar. The net effect of these changes has an overall positive impact on the total error rate from the \texttt{NON-GAMIFIED} set-up to the \texttt{GAMIFIED-UNLIMITED} set-up: from $41.52\% \rightarrow 37.16\%$ (level 2) and from $35.51\% \rightarrow 31.48\%$ (level 3). One possible interpretation is that the feedback helps listeners to be aware of their high miss rate and consequently level miss and false alarm rates closer to one another. As a heuristic measure, the last column in Table \ref{tab:error-rates-by-level} shows the ratio of the average miss and false alarm rates, $P_\text{miss}/P_\text{fa}$. A value 1.0 would indicate that the two types of errors are in perfect balance (analogous to equal error rate operating point in automatic systems whereby $P_\text{miss}=P_\text{fa}$). For levels 2 and 3 these ratios are indeed closer to 1.0 in both gamified designs compared to non-gamified design.

Any comparison with \texttt{GAMIFIED-LIMITED} must consider that the number of workers that could perform levels 2 and 3 is smaller than in \texttt{NON-GAMIFIED} set-up (22 workers completed level 3 in \texttt{NON-GAMIFIED} while only 9 workers answered at least one trial in level 3 for \texttt{GAMIFIED-LIMITED}). Though the total error rate is slightly higher in level 2 and larger in level 3, the miss and false alarm rates for workers in \texttt{GAMIFIED-LIMITED} are similar for level 3 while in \texttt{NON-GAMIFIED} the miss rates are six times the false alarm. This suggests that in the gamified set-up with limited 'lives' workers try to balance their performance for target and nontarget trials. 

\begin{figure}[!t]
\centering
\includegraphics[width=.4\textwidth]{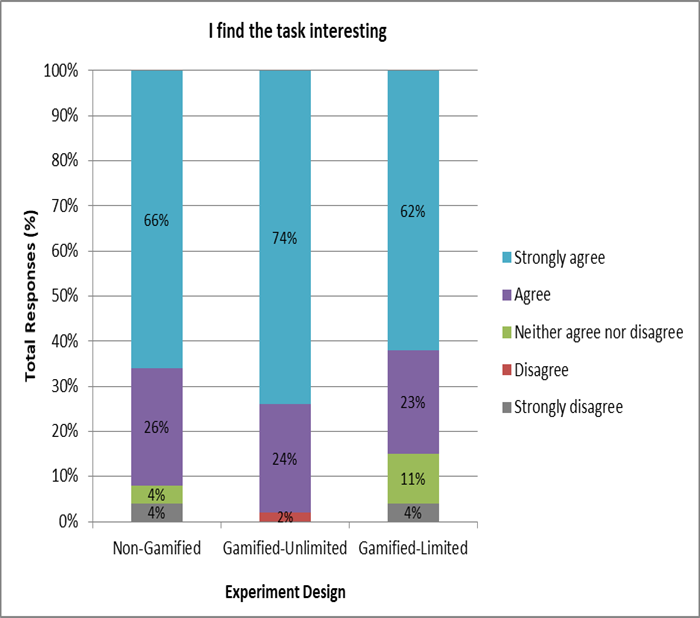}
\caption{Survey on listener's interest for performing the task}
\label{fig:interest_survey}
\includegraphics[width=.4\textwidth]{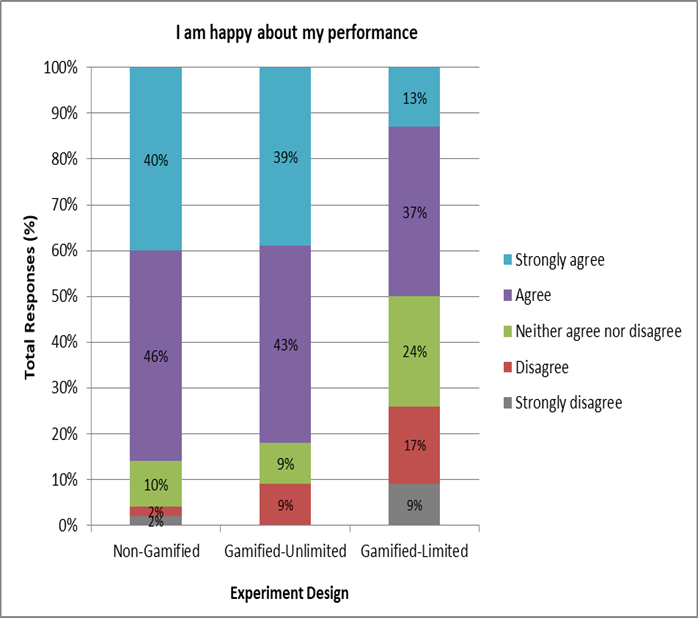}
\caption{Survey on satisfaction of listeners about their performance}
\label{fig:satisfaction_survey}
\includegraphics[width=.4\textwidth]{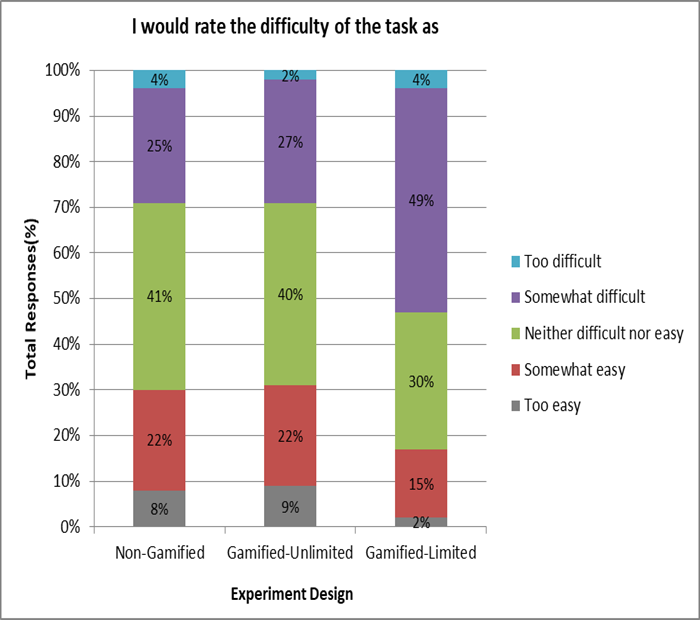}
\caption{Survey on perceived difficulty of the task by listeners}
\label{fig:difficulty_survey}
\end{figure}


\subsection{Quitting time}


We now turn our attention to the time instant (trial id) at which the workers quits the experiment, visualized in Figure \ref{fig:quit_chart} for the three interface designs. Comparing \texttt{NON-GAMIFIED} and \texttt{GAMIFIED-UNLIMITED}, both follow a common pattern where a random worker quits with high probability \emph{either} after completing the mandatory part, 10 trials of level 1, \emph{or} after completing all the 30 trials. For both set-ups, a good portion of the workers were motivated to continue until the end, and a minority of the workers quitting in between. 

The \texttt{GAMIFIED-LIMITED} set-up with limited number of lives, however, is different. While the two largest modes now occur after completing either one of the first two levels, there are \emph{no} workers who reach the very end\footnote{The authors felt highly apologetic for the two workers who reached trial 29/30; despite not meeting our formal criteria (30), we granted them the bonus (0.50 USD).}. The sparsifying trend towards the last game level is readily understood by observing the difference between players who decided to quit voluntarily vs. who ran out of `lives' (See Fig. \ref{fig:quit_chart}c).




\subsection{Probability Calibration}

Our next analysis concerns relation between a worker's performance (accuracy) vs. the self-reported confidence values (a number between 1-100 given for each trial). Intuitively, discrepancy between the two can be used to assess \emph{probability calibration}. 
Consider a hypothetical example. A worker who completes all the 30 trials (15 target and 15 nontarget) and reports the same confidence levels of 60\% (=10/15) and 80\% (=12/15) for target and nontarget trials, respectively. Such worker is \emph{perfectly calibrated} if the corresponding actual accuracy for targets and nontargets aligns with these numbers. A worker who reports higher confidence values than his or her actual accuracy is \emph{overconfident}, while the opposite case represents \emph{underconfidence}.

In Figure \ref{fig:confidence_plot}, we visualize the success rate of each worker against their averaged self-reported confidence for target and non-target trials\footnote{The Pearson correlation ($r$) and $p$-values for self-reported confidence and success for target and non-target trials respectively are: \texttt{NON-GAMIFIED}: $r=-0.087, p=0.55$, $r=0.48, p<0.005$; \texttt{GAMIFIED-UNL}: $r=0.22, p=0.13$, $r=0.15, p=0.31$ and \texttt{GAMIFIED-LIM}: $r=0.12, p=0.41$, $r=0.29, p=0.043$.}. The diagonal dotted line represents the ideal perfectly calibrated cases, whereas the points above and below diagonal represent overconfidence and underconfidence respectively. In all three interface designs, most of the workers are observed to be overconfident on target trials and underconfident on non-target trials. Interestingly, a dense set of points for high success rate (100\%) is observed for non-target trials compared to target trials. It indicates that workers were better at identifying 'different speakers' though they were more confident on 'same speakers' trials.


\subsection{Survey/self-assessment analysis}
In order to compare the worker's interest, satisfaction and perception towards three different interface designs, we conducted a short survey at the end of the game. The workers were asked to rate the following statements on a 5-point rating scale: 
\begin{enumerate}
\item \emph{I find the task interesting (Options: Strongly agree, Agree, Neither agree nor disagree, Disagree, Strongly disagree)}
\item \emph{I am happy about my performance (Options: Strongly agree, Agree, Neither agree nor disagree, Disagree, Strongly disagree)}
\item \emph{I would rate the difficulty of the task as (Options: Too difficult, Somewhat difficult, Neither difficult nor easy, Somewhat easy, Too easy)}
\end{enumerate}

Figures \ref{fig:interest_survey}, \ref{fig:satisfaction_survey} and \ref{fig:difficulty_survey}, visualize the results comparing each interface design. In Figure \ref{fig:interest_survey}, majority of workers claim that they find the task interesting in all three scenarios which indicates positive appeal towards speaker comparison tasks in general by human listeners. Figure \ref{fig:satisfaction_survey} shows interesting variation on worker's satisfaction in \texttt{GAMIFIED-UNLIMITED} and \texttt{GAMIFIED-LIMITED} interfaces. In \texttt{GAMIFIED-LIMITED} scenario, the reason for less number of satisfied workers than \texttt{GAMIFIED-UNLIMITED} could be due to the fact that most of them were not able to complete the game or reach the highest level. The same trend persists for rating the difficulty of the task as well possibly for the same reason.

\section{Conclusion}
We addressed speaker comparison by listening through a game-like interface design. In reference to the questions posed in Section 1, our data reveals the following.
    \begin{enumerate}
        \item Our answer is partially affirmative. With reference to Table \ref{tab:average_error_rates}, we saw a decrease in the average total error rate from non-gamified setup to the gamified setup with unlimited lives. For the gamified design with limited lives, however, this did not hold. 
        \item Concerning the two types of errors, the number of misses was observed to be significantly higher than the number of false alarms, regardless of the interface design. This might relate to the type of material in VoxCeleb. An interesting finding, however, was that the average miss rates were substantially lowered in the two gamified designs. As a result, the two types of errors were found more balanced in the gamified designs. We suspect that the feedback provided in the gamified designs helped listeners to adjust their decisions. 
        \item Interestingly, many listeners were systematically \emph{overconfident on target trials} and systematically \emph{underconfident on nontarget trials}, regardless of the interface design. Overall, our listeners were far from being well-calibrated. 
        \item Comparing the non-gamified design with the unlimited gamified design, the listeners found the latter more interesting (non-gamified: 92\% rated `agree' or `strongly agree'; gamified: 98\%). The gamified set-up with limited lives, however, was perceived as less interesting (85\%). Self-satisfaction and perceived easiness were also clearly lower in the limited design compared to the other two. These findings are in reasonable agreement with the objective performance.
    \end{enumerate}

Overall, we find the preliminary results encouraging, indicating the potential of gamified speaker comparison designs. In principle, intrinsically-motivated crowdworkers could perform better, and our data has preliminary indications of this. 

We also acknowledge the limitations of our study. Due to both budget constraints and inherent listener limitations, it is infeasible to carry out experiments with many different interface designs or large number of trials. This necessarily limits the statistical power of findings. We nonetheless hope our work to inspire further research or applications. A game perceived sufficiently `cool' may have the potential to go viral. It may not be unrealistic to have massive crowd of listeners to take part in validating speaker labels in massive \emph{found data} collections whose number keeps increasing. 



\bibliographystyle{IEEEbib}
\bibliography{Odyssey2022_BibEntries}

%

\end{document}